\documentclass[aip,jcp,reprint,noshowkeys]{revtex4-1}
\usepackage{graphicx,dcolumn,bm,xcolor,microtype,multirow,amscd,amsmath,amssymb,amsfonts,physics,mhchem}

\usepackage{mathpazo,libertine}
\usepackage[normalem]{ulem}
\definecolor{darkgreen}{rgb}{0.13, 0.55, 0.13}
\newcommand{\alter}[1]{\textcolor{black}{#1}}

\usepackage{hyperref}
\hypersetup{
    colorlinks=true,
    linkcolor=blue,
    filecolor=blue,      
    urlcolor=blue,	
    citecolor=blue
}

\newcommand{\br}{\boldsymbol{r}}
\newcommand{\hH}{\Hat{H}_{\lambda}}
\newcommand{\hT}{\Hat{T}}

\newcommand{\PsiUHF}{\Psi_\text{UHF}}

\newcommand{\ERHF}{\widetilde{E}_\text{RHF}}
\newcommand{\EUHF}{\widetilde{E}_\text{UHF}}

\newcommand{\LCPQ}{Laboratoire de Chimie et Physique Quantiques (UMR 5626), Universit\'e de Toulouse, CNRS, UPS, France}
\newcommand{\UCAM}{Department of Chemistry, University of Cambridge, Lensfield Road, Cambridge, CB2 1EW, U.K.}

\begin{document}	

\title{Complex Adiabatic Connection: a Hidden Non-Hermitian Path from Ground to Excited States}

\author{Hugh G.~A.~\surname{Burton}}
\thanks{Corresponding author}
\email{hb407@cam.ac.uk}
\affiliation{\UCAM}
\author{Alex J.~W.~\surname{Thom}}
\email{ajwt3@cam.ac.uk}
\affiliation{\UCAM}
\author{Pierre-Fran\c{c}ois \surname{Loos}}
\email{loos@irsamc.ups-tlse.fr}
\affiliation{\LCPQ}

\begin{abstract}
Processes related to electronically excited states are central in many areas of science, \alter{however accurately determining excited-state energies remains a major challenge in theoretical chemistry.
Recently, higher energy stationary states of non-linear methods have themselves been proposed as approximations to excited states, although the general understanding of the nature of these solutions remains surprisingly limited.}
In this Letter, we present an entirely novel approach for \alter{exploring and} obtaining excited \alter{stationary} states by exploiting the properties of non-Hermitian Hamiltonians. 
Our key idea centres on performing \alter{analytic continuations of} conventional \alter{quantum} chemistry methods.
Considering Hartree--Fock theory as an example, we \alter{analytically continue the electron-electron interaction} to expose a hidden connectivity of multiple solutions \alter{across the complex plane}, revealing a close resemblance between Coulson--Fischer points and non-Hermitian degeneracies.
Finally, we demonstrate how a ground-state wave function can be morphed naturally into an excited-state wave function \alter{by constructing a well-defined complex adiabatic connection.}
\end{abstract}

\maketitle

\textit{Introduction.---}
Electronic excited states are central in chemistry, physics and biology, \alter{playing a role in key} processes such as photochemistry, catalysis, and solar cell technology.
\alter{However, defining effective methods} that reliably provide accurate excited-state energies remains a major challenge in theoretical chemistry.
Two of the most widely-used approaches to obtain excited-state energies are i) the time-dependent (TD) version of density-functional theory (DFT) which relies on the linear response formalism, and ii) the equation-of-motion (EOM) \textit{ansatz} of coupled cluster (CC) theory.

In particular, TD-DFT has practically revolutionised computational chemistry due to its user-friendly black-box nature compared with the more computationally expensive multi-configurational methods (such as CASPT2 and NEVPT2) where one must choose an active space based on chemical intuition.
\alter{Despite their success}, fundamental deficiencies associated with TD-DFT and EOM-CC remain. 
For example, excited states presenting double excitation character \cite{Elliott_2011, Maitra_2004, Cave_2004, Maitra_2012, Krylov_2001, Sagredo_2018, Loos_2018b, Loos_2019} \alter{--- which have a key role in the faithful description of many physical and chemical processes ---} are notoriously difficult to model using conventional single-reference methods such as adiabatic TD-DFT or EOM-CC. 
Although some viable and promising alternative approaches have been developed --- for example spin-flip, \cite{Krylov_2001} dressed TD-DFT \cite{Maitra_2004} or ensemble DFT \cite{Sagredo_2018} --- each faces major limitations.

\alter{
At present, most excited-state techniques, including TD-DFT and EOM-CC, are built upon a single reference Slater determinant, often corresponding to a Hartree--Fock (HF) solution.
As an inherently non-linear method, similar to CC \cite{Kowalski_1998} and GW, \cite{Kozik_2014, Stan_2015, Rossi_2015, Tarantino_2017, Schaefer_2013, Schaefer_2016, Gunnarsson_2017, Veril_2018} HF can produce a multitude of distinct stationary states.
In recent years, multiple HF states have themselves been proposed as approximations to excited states.\cite{Gilbert_2008, Barca_2014, Peng_2013, Jensen_2018}
However, these solutions do not necessarily share the symmetries of the exact Hamiltonian,\cite{Lykos_1963,StuberPaldus} and the onset of symmetry breaking, where multiple solutions coalesce at so-called ``Coulson--Fischer points'',\cite{Coulson_1949} appears intimately linked to the strength of the electron-electron (e-e) interaction present.\cite{Jiminez-Hoyos_2011, Lee_2018}
For cases where the single-determinant representation fails, the utilisation of these multiple HF solutions as a basis for non-orthogonal configuration interaction (NOCI) has been shown to recover symmetry-pure multi-reference ground and excited state energies.\cite{Ayala_1998, Thom_2009, Sundstrom_2014, Mayhall_2014, Burton_2016, Jensen_2018}
Despite significant progress, however, our understanding of the general nature of multiple solutions remains surprisingly limited. \cite{Hiscock_2014, Burton_2016,  Burton_2018, Thouless_1960, Cizek_1967, Paldus_1969, Seeger_1977, Fukutome_1981, StuberPaldus, Thom_2008, Gilbert_2008,  Barca_2014, Barca_2018a, Barca_2018b}
}

\alter{In this Letter, we propose a totally novel approach for exploring multiple solutions in electronic structure methods.
Our key idea focuses on performing complex analytic continuations of conventional methods which, by exploiting the properties of non-Hermitian Hamiltonians, reveal hidden features of multiple stationary states.
Although more sophisticated methods will be considered in the future (in particular DFT, CC and GW), we illustrate the general approach by introducing a complex-scaled e-e interaction in the HF method.
In doing so, we expose a deeper topology of connected stationary states across the complex plane, and identify a close resemblance between Coulson--Fischer points and non-Hermitian degeneracies.
Finally we demonstrate how, through this complex landscape, ground and excited states can be naturally interconverted \textit{via} a complex adiabatic connection.
}

\textit{Non-Hermitian quantum mechanics.---}
Our understanding of quantum systems has been transformed by the introduction of non-Hermitian Hamiltonians\cite{Bender_1998} --- a complex generalisation of conventional quantum mechanics \cite{Bender_1998, Bender_2002} --- as an approach for exploring multiple eigenstates through the framework of complex analytic continuation.
Using this technique, a real-symmetric Hamiltonian is analytically continued into the complex plane, becoming non-Hermitian in the process and exposing the fundamental topology of eigenstates. 
For example, one of the most amazing aspects of non-Hermitian quantum mechanics is that quantised eigenvalues emerge directly from the different sheets of a Riemann surface.\cite{Bender_2008}
In other words, our view of the quantised nature of conventional Hermitian quantum mechanics arises only from our limited perception of the more complex and profound structure of its non-Hermitian variant.

To our knowledge, the multiple solutions to the non-linear HF equations remain unexplored in the framework of  analytic continuation.
Since the conventional complex Hermitian extension of HF theory violates the Cauchy--Riemann conditions (resulting in functions that are not complex analytic), we rely here on the holomorphic HF (h-HF) approach \cite{Hiscock_2014, Burton_2016, Burton_2018} originally developed as a method for analytically continuing real HF solutions beyond the Coulson--Fischer points at which they coalesce and vanish.\cite{Mestechkin_1979, Mestechkin_1988}
In h-HF theory, the complex conjugation of orbital coefficients is simply removed from the conventional HF equations, resulting in a non-Hermitian Hamiltonian and an energy function that is complex analytic with respect to the orbital coefficients. 
When the orbital coefficients are real, the HF and h-HF formalisms are equivalent. 
However, h-HF solutions are found to exist over the full potential energy surface, obtaining complex orbital coefficients when their real counterparts coalesce and disappear.\cite{Hiscock_2014, Burton_2016, Burton_2018}

\begin{figure}
	\includegraphics[width=0.8\linewidth]{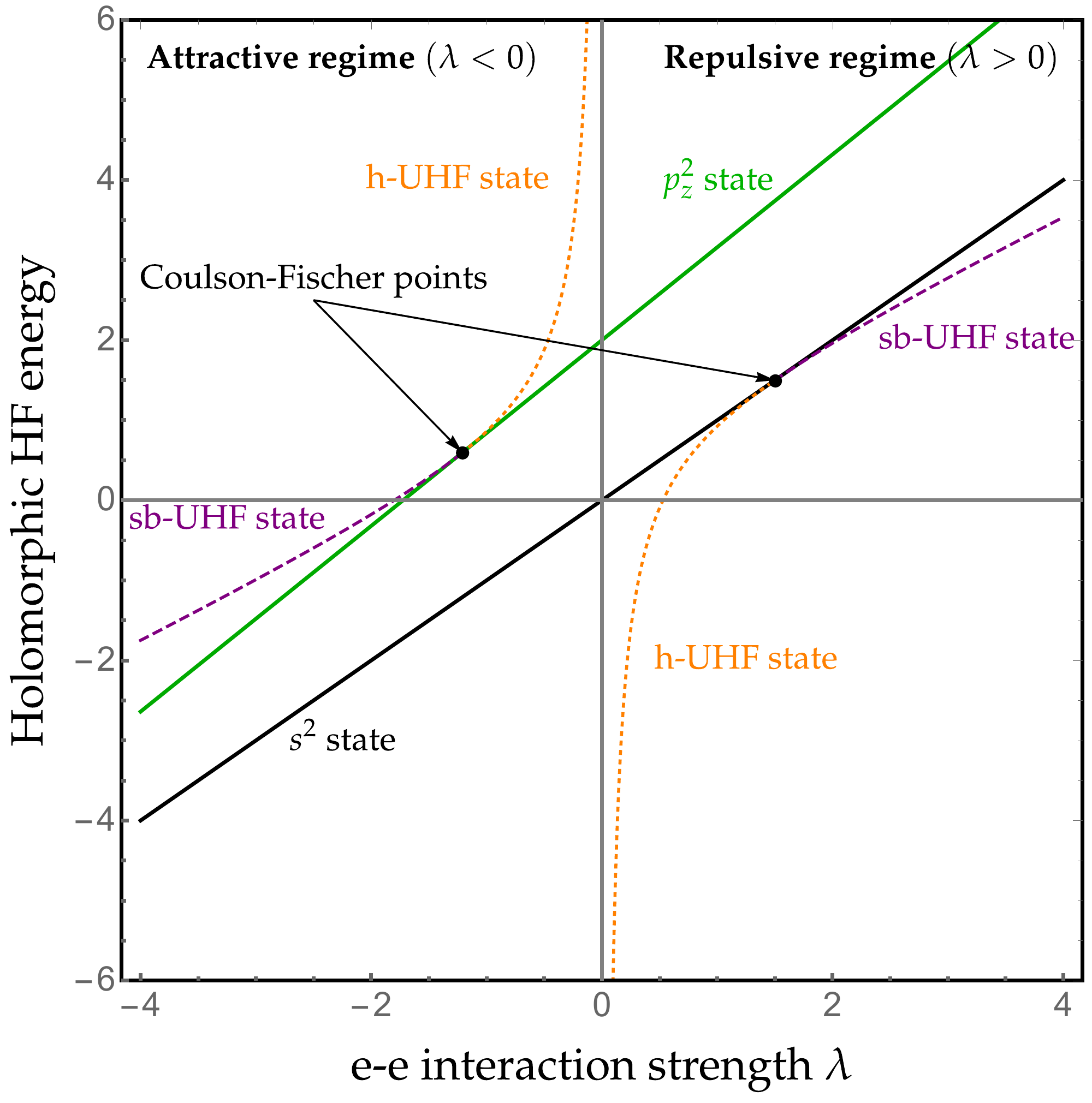}
	\caption{
	\label{fig:energy}
	Holomorphic energy of the different HF solutions as functions of the e-e interaction strength $\lambda$.
	The holomorphic energy of the symmetry-broken h-UHF states becomes singular at $\lambda = 0$.
	}
\end{figure}

The use of non-Hermitian Hamiltonians in quantum chemistry is not itself new; these Hamiltonians have been used extensively as a method for describing metastable resonance phenomena. \cite{MoiseyevBook}
Through a complex-scaling of the electronic or atomic coordinates,\cite{Moiseyev_1998} or by introducing a complex absorbing potential,\cite{Riss_1993, Ernzerhof_2006} \alter{outgoing} resonance states are transformed into square-integrable wave functions that allow the energy and lifetime of the resonance to be computed (see Ref.~\onlinecite{MoiseyevBook} for a general overview).

Although Hermitian and non-Hermitian Hamiltonians are closely related, the behaviour of their eigenvalues near degeneracies is starkly different.\cite{Feuerbacher_2004, Benda_2018}
For example encircling non-Hermitian degeneracies at ``exceptional points'' leads to the interconversion of states\cite{Feuerbacher_2004, Hernandez_2006} and can apply a geometric phase.\cite{Heiss_1999}
In contrast, encircling Hermitian degeneracies at ``conical intersections'' \cite{Yarkony_1996} introduces only a geometric phase, leaving the states unchanged.\cite{Berry_1984}
More dramatically, whilst eigenvectors remain orthogonal at conical intersections, at non-Hermitian exceptional points the eigenvectors themselves become equivalent.
The result is a self-orthogonal state and a set of eigenvectors that no longer span the full space.\cite{MoiseyevBook}

The fascinating aspect of non-Hermitian Hamiltonians --- and the launch-pad for the remainder of this Letter --- is the claim that, by analytically continuing the e-e interaction, an underlying landscape of solutions can be revealed in which a ground-state wave function can be morphed into an excited-state wave function by following a well-defined contour in the complex plane.
In what follows, atomic units are used throughout.

\begin{figure*}
	\includegraphics[width=0.33\linewidth]{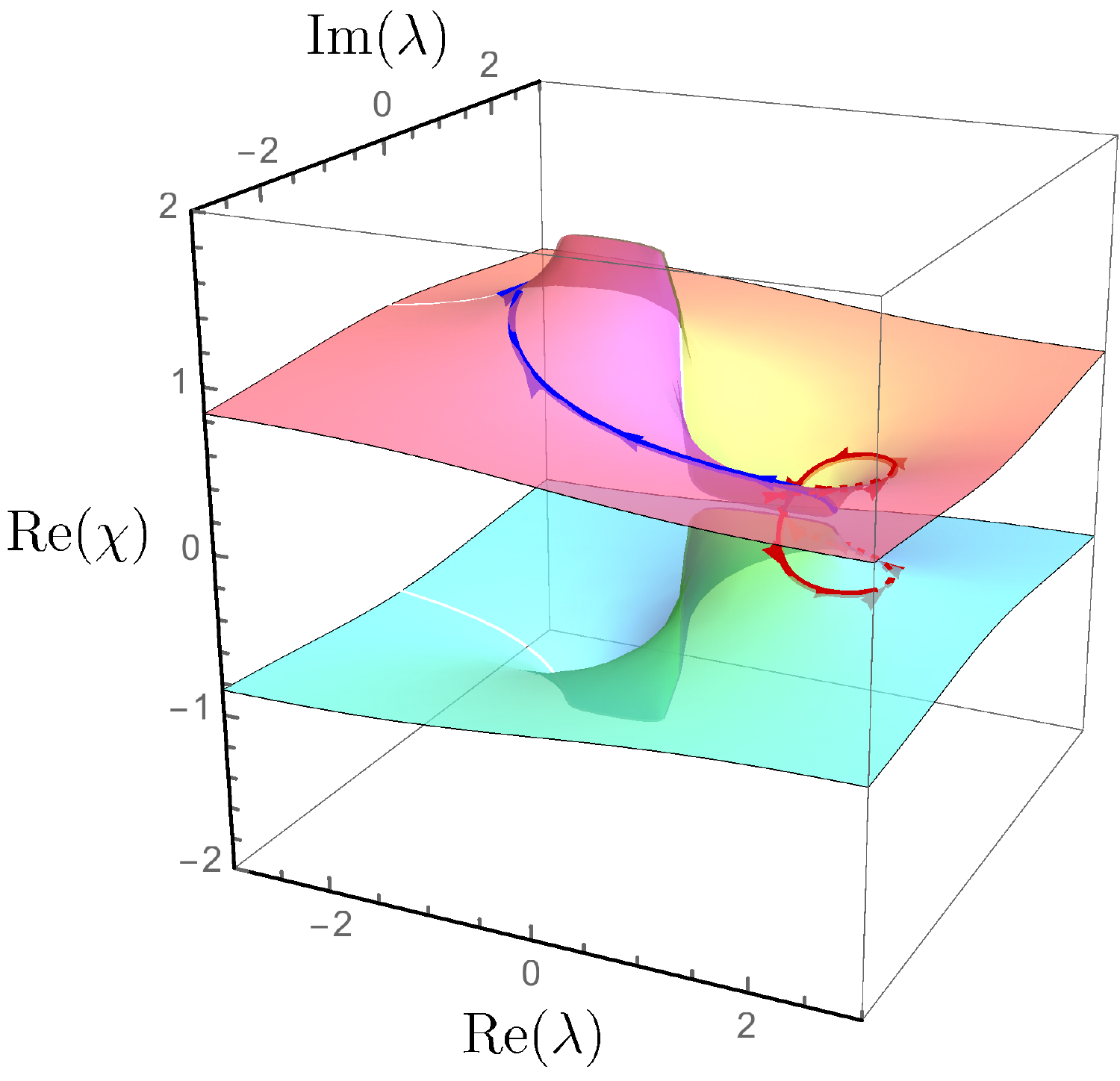}
	\includegraphics[width=0.33\linewidth]{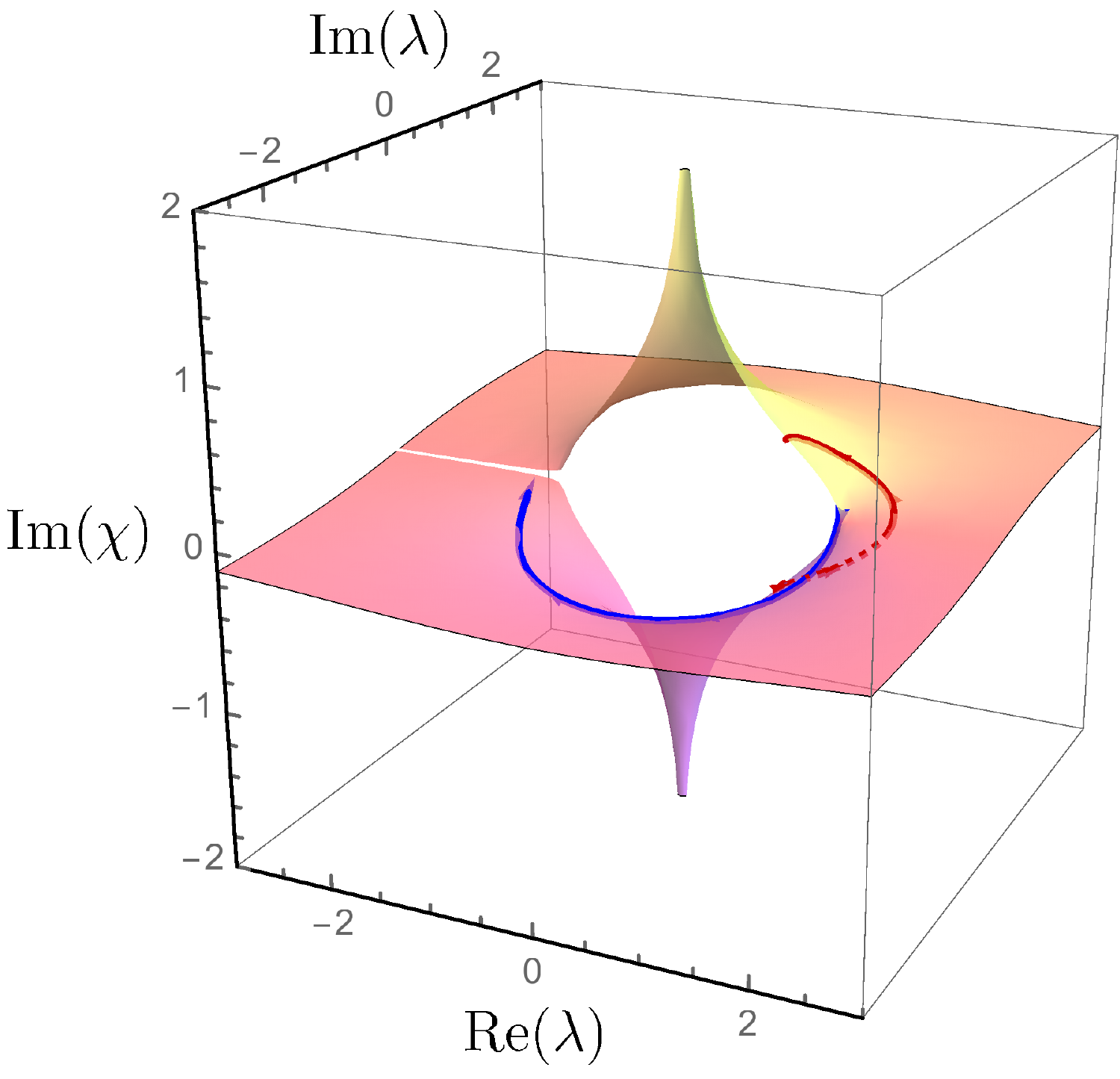}
	\includegraphics[width=0.33\linewidth]{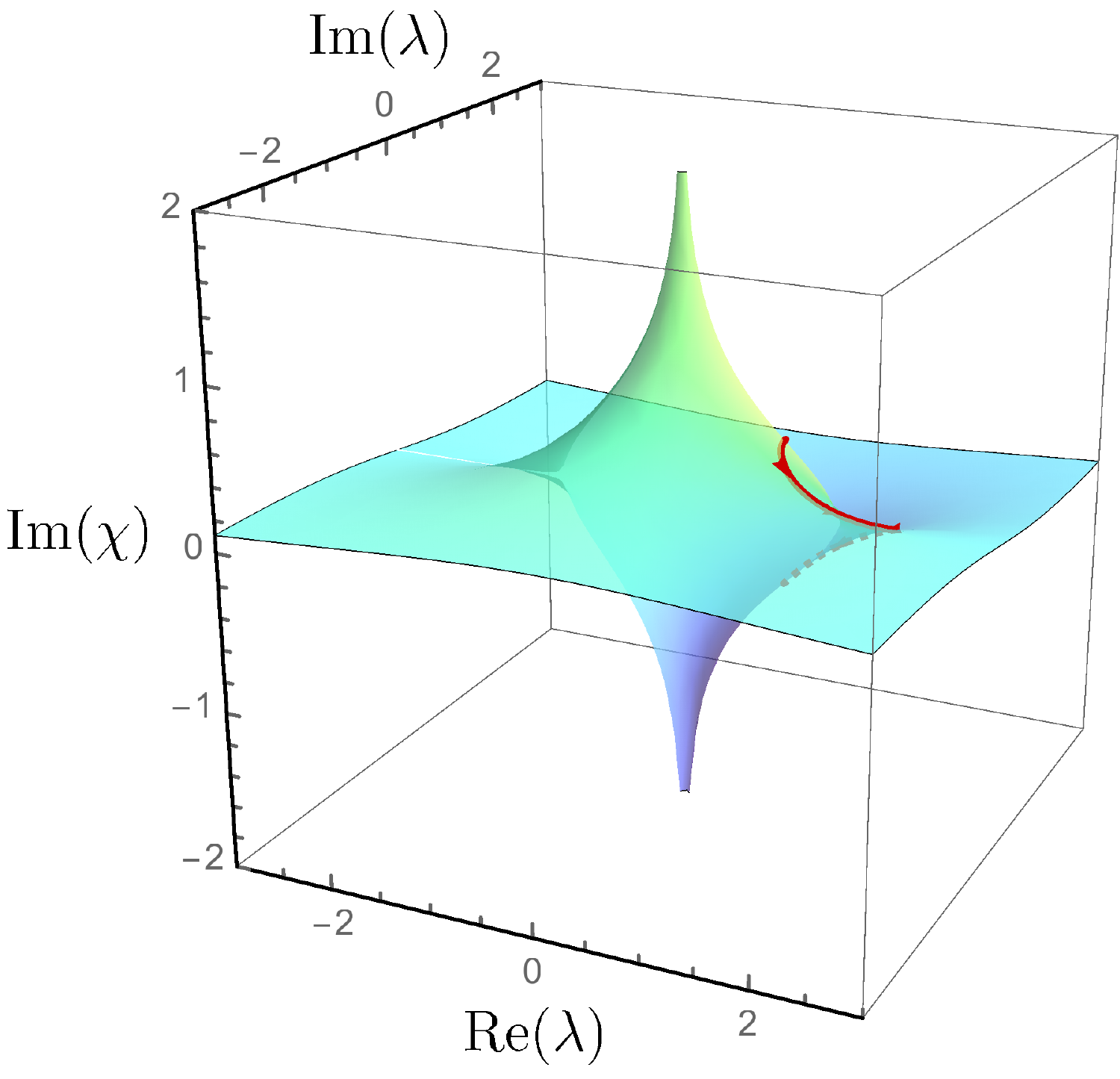}
	\\
	\includegraphics[width=0.25\linewidth]{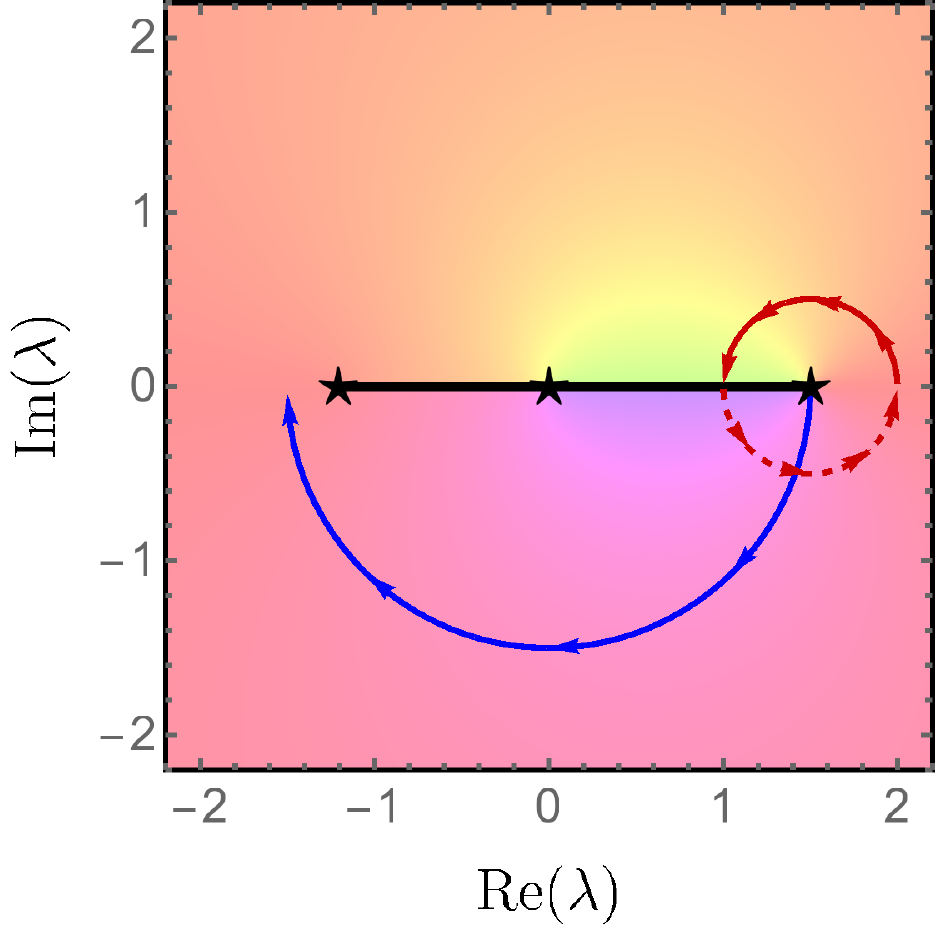}
	\hspace{0.1\linewidth}
	\includegraphics[width=0.25\linewidth]{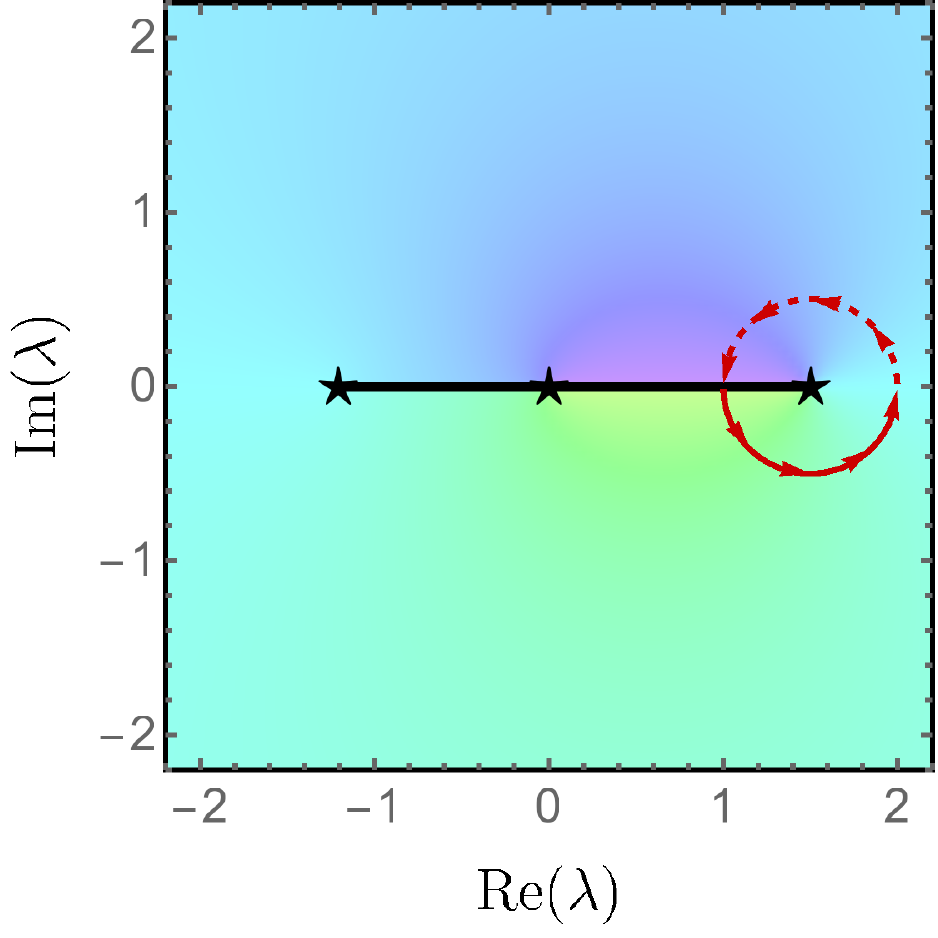}
\caption{
	Real (top left) and imaginary (top middle and top right) components of the mixing angle $\chi$ for the two degenerate sb-UHF solutions, Eq.~\eqref{eq:chi_sbUHF}, as a function of the real and imaginary parts of $\lambda$.
    The colouring indicates the phase of $\chi$.
	Periodic repeats of these surfaces exist for $\Re ( \chi ) < -\pi/2$ and $\Re ( \chi ) > \pi/2$ representing equivalent states to those shown.
    The RHF solutions are given by planes at $\chi = 0$ and $\pi/2$, however these are omitted for clarity.
	The two sheets of the Riemann surface (bottom left and bottom right) show three branch points at $\lambda = -75/62$, $0$ and $3/2$ (black stars), connected by a branch cut (solid black).
	Following a path once around the branch point at $\lambda = 3/2$ interconverts the two sb-UHF solutions (solid red), whilst completing a second rotation returns the solutions to their original states (dashed red).
	The adiabatic contour enabling a smooth transition from the ground state to the excited state is represented by the solid blue curve.
	\label{fig:riemann}}
\end{figure*}

\textit{A simple model.---}
Let us illustrate the general idea of the present Letter by considering a very simple model system comprising two opposite-spin electrons interacting through the long-range Coulomb potential whilst confined to the surface of a sphere of radius $R$, which we set to unity for convenience. Moreover, let us consider only two basis functions: an $s$-type orbital [$s \equiv Y_{0}(\theta)$] and a $p$-type orbital [$p_z \equiv Y_{1}(\theta)$], where $Y_\ell(\theta)$ are zonal harmonics, \cite{NISTbook} and $\theta$ is the polar angle of the electron.
To control the strength of the e-e interaction, we introduce the adiabatic scaling parameter $\lambda$, giving $\lambda = 0$ for the non-interacting system and $\lambda = 1$ for the physical (i.e.~interacting) system.
The e-e scaled Hamiltonian is therefore given by
\begin{equation}
\label{eq:H}
    \hH = \hT + \lambda\,r_{12}^{-1},
\end{equation}
where $\hT = -(\nabla_1^2 + \nabla_2^2$)/2 is the combined kinetic operator for the two electrons, and $r_{12}^{-1} \equiv \abs{\br_1 - \br_2}^{-1}$ is the Coulomb operator. Note that the two electrons interact \textit{through} the sphere.
The ``two-electrons-on-a-sphere'' paradigm (see Ref.~\onlinecite{Loos_2009a} for more details) possesses a number of interesting features, \cite{Loos_2009a, Loos_2009c, Loos_2011b, Loos_2018a} and it can be seen as a unique theoretical laboratory to test various theoretical methods. \cite{Loos_2018a}

Here, we wish to illustrate how one can obtain the restricted HF (RHF) doubly-excited state $p_z^2$ starting from the RHF ground-state $s^2$ configuration, a process which is not as easy as one might think, and particularly challenging with conventional self-consistent field algorithms. \cite{Thom_2008, Gilbert_2008, Shea_2018}
Similar to the \ce{H2} molecule (see Ref.~\onlinecite{SzaboBook} for a pedagogical discussion), we define an unrestricted HF (UHF) wave function
\begin{equation}
	\PsiUHF(\theta_1,\theta_2) = \varphi(\theta_1) \varphi(\pi - \theta_2),
\end{equation}
where the spatial orbital is 
\begin{equation}
	\varphi =  s \cos \chi +  p_z \sin \chi,
\end{equation}
$\chi$ is the mixing angle between the two basis functions, and the associated holomorphic energy is 
\begin{equation}
\begin{split}
	\EUHF(\chi,\lambda) & = \qty(1-\cos 2\chi) 
	\\
						& + \frac{\lambda}{75} \qty(67-6\cos 2\chi+14\cos 4\chi).
\end{split}
\end{equation}
Ensuring the stationarity of the UHF energy, i.e.,~$\pdv*{\EUHF}{\chi} = 0$, one obtains
\begin{equation}
\label{eq:dEUHF}
    \sin 2 \chi \left( 75 + 6 \lambda - 56 \lambda \cos 2\chi \right) = 0.
\end{equation}
For $\chi = 0$ and $\pi/2$, we recover the RHF $s^2$ ground state and the $p_z^2$ doubly-excited state with respective holomorphic energies 
\begin{align}
\label{eq:RHF}
	\ERHF^{s^2}(\lambda) & = \lambda,
	&
	\ERHF^{p_z^2}(\lambda) & = 2 + \frac{29 \lambda}{25}.
\end{align}
These are represented, as a function of $\lambda$, by the black and green solid lines in Fig.~\ref{fig:energy}. 
The two-fold degenerate UHF solutions (mutually related by spin-flip symmetry) are given by
\begin{equation}
\label{eq:chi_sbUHF}
    2\chi = \pm \arccos\qty( \frac{3}{28} + \frac{75}{56 \lambda} ),
\end{equation}
with holomorphic energy
\begin{equation}
\label{eq:EUHF}
	\EUHF(\lambda) = - \frac{75}{112 \lambda} +  \frac{25}{28} + \frac{59 \lambda}{84}.
\end{equation}

For $\lambda > 3/2$, the UHF wave function is a real-valued ``symmetry-broken'' UHF (sb-UHF) solution of the ground-state RHF wave function, while it is a real-valued sb-UHF solution of the excited RHF wave function for $\lambda < -75/62$ (purple dashed lines in Fig.~\ref{fig:energy}).
For $-75/62 < \lambda < 3/2$, the UHF solution is a holomorphic UHF (h-UHF) solution with complex coefficients (orange dotted lines in Fig.~\ref{fig:energy}).
Its holomorphic energy, though, still given by Eq.~\eqref{eq:EUHF}, stays real.
These energies are represented  as functions of $\lambda$ in Fig.~\ref{fig:energy}, where one can observe two distinct regimes: the repulsive regime ($\lambda > 0$) and the attractive regime ($\lambda < 0$).
The Coulson--Fischer points (black dots in Fig.~\ref{fig:energy}) correspond to the $\lambda$ values where the RHF and sb-UHF solutions coalesce, and are located at the ``kissing'' points of Eq.~\eqref{eq:EUHF} with the ground and excited RHF states [see Eq.~\eqref{eq:RHF}].

\begin{figure*}
	\includegraphics[height=0.36\linewidth]{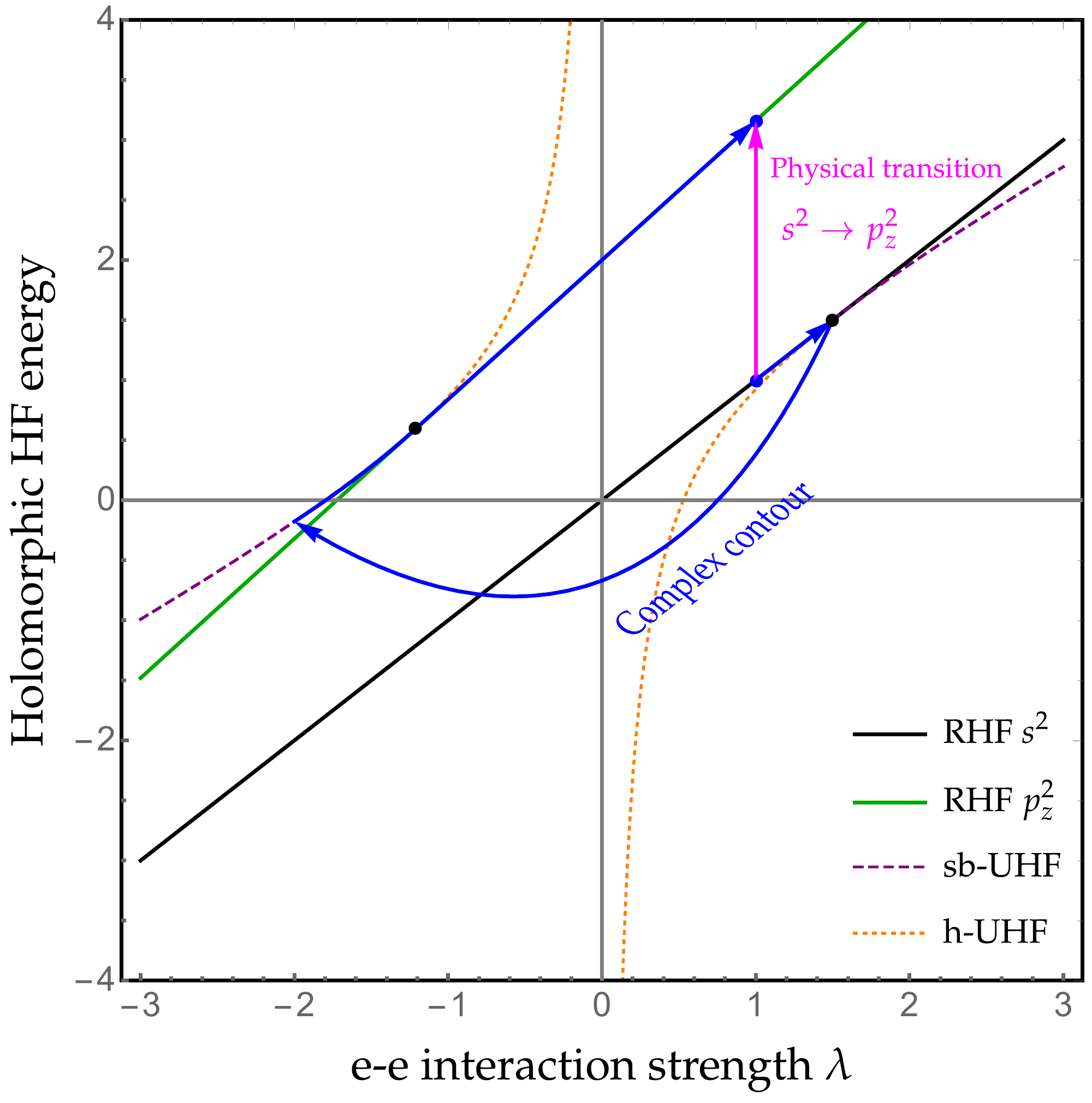}
    \hspace{0.05\linewidth}
	\includegraphics[height=0.38\linewidth]{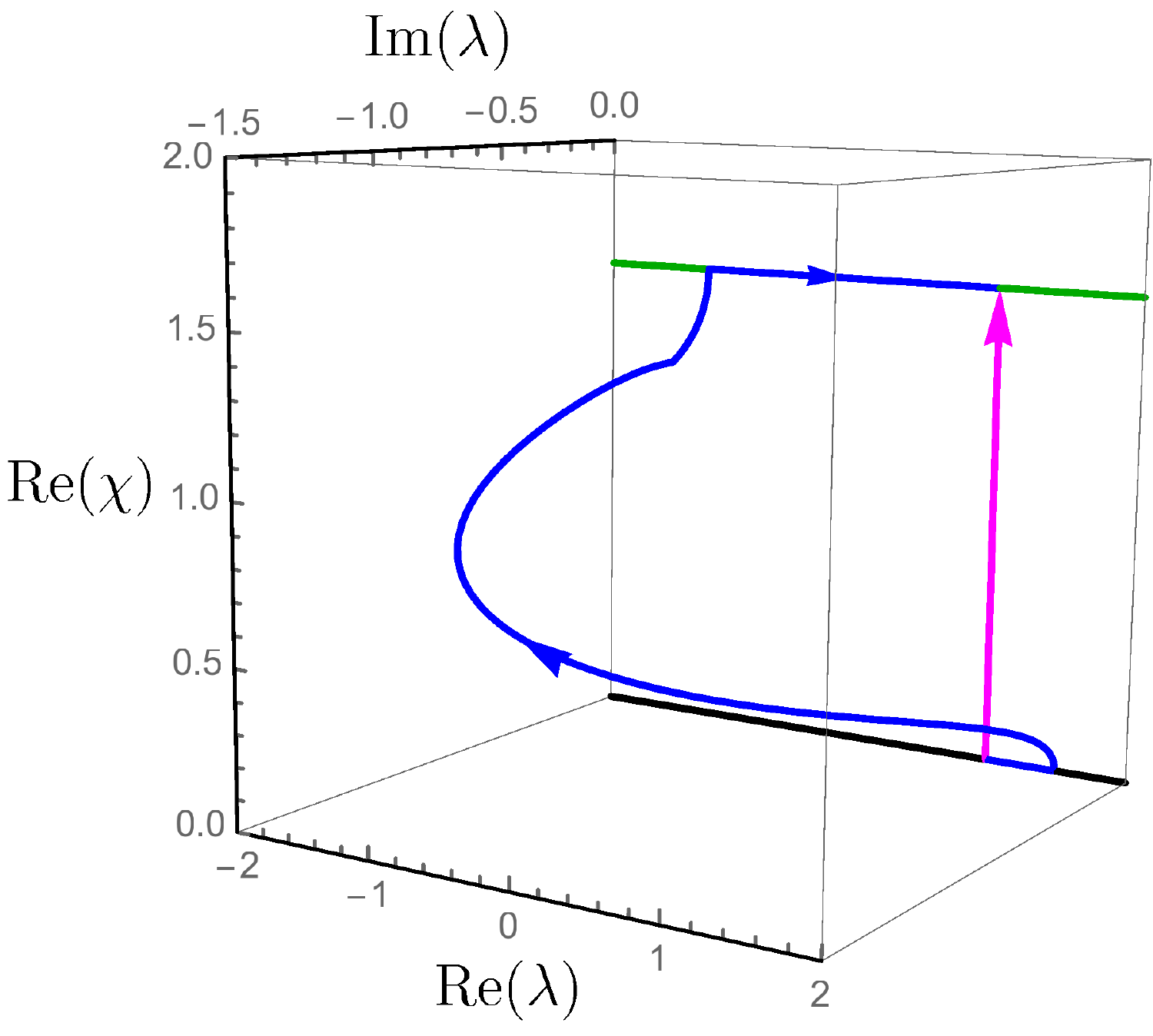}
	\caption{
	\label{fig:adia}
	An example of a complex adiabatic connection path (in blue) enabling the physical transition $s^2 \rightarrow p_z^2$ (at $\lambda = 1$) to be obtained.
	The contour followed in the complex plane is represented diagrammatically against the holomorphic energies (left), whilst the real component of $\chi$ along the contour demonstrates the connection pathway from the ground to the excited state (right). 
	}
\end{figure*}
\textit{Quasi-exceptional points.---}
The description above provides only a glimpse into the fundamental nature of sb-UHF solutions.
By analytically extending $\lambda$ into the complex plane, i.e.~taking
\begin{equation}
	\arccos(z) = \pi/2 + \mathrm{i} \log( \mathrm{i} \,z+\sqrt{1-z^2}),
\end{equation}
where $z = 3/28 + 75/(56\lambda)$, 
the solutions to Eq.~\eqref{eq:chi_sbUHF} elegantly emerge as two sheets of a Riemann surface, shown in Fig.~\ref{fig:riemann}. 
From this surface, we see that the apparent discrete nature of multiple HF solutions arises by considering only the real $\lambda$-axis, and that sb-UHF states are in fact part of a wider unified structure.
The Coulson--Fischer points in our more general picture correspond to branch points at $\lambda = 3/2$ and $-75/62$ (black stars in Fig.~\ref{fig:riemann}), connected by a branch cut (red line in Fig.~\ref{fig:riemann}). 
By following a path around one of these points, for example at $\lambda = 3/2$, we observe the interconversion of the two coalescing sb-UHF states (red solid lines in Fig.~\ref{fig:riemann}). 
Completing a second rotation restores the solutions to their original states, although no geometric phase occurs (red dashed lines in Fig.~\ref{fig:riemann}).
Significantly, by extending $\lambda$ into the complex plane we have revealed that Coulson--Fischer points behave more generally as exceptional points of the non-Hermitian e-e~scaled Hamiltonian, despite the absence of any resonances or continuum in our closed system.

This correspondence between Coulson--Fischer points and conventional exceptional points is not strictly exact. 
Due to the non-linearity of the HF equations, its multiple solutions need not be mutually orthogonal and we find that, although the sb-UHF wave functions coalesce at the Coulson--Fischer point, they do not become self-orthogonal.\cite{MoiseyevBook}
We believe this is the first reported occurrence of exceptional points without self-orthogonality, and we henceforth refer to such peculiar phenomena as ``quasi-exceptional points''.
In contrast, the additional logarithmic branch point that appears at $\lambda = 0$ \textit{does} behave as a genuine exceptional point.
Here, the self-consistent e-e interaction is removed completely from the HF equations [see Eq.~\eqref{eq:H}] and the solutions must share the spherical symmetry of the non-interacting kinetic operator.
For the sb-UHF solutions in this limit, the mixing angle [see Eq.~\eqref{eq:chi_sbUHF}] becomes $\chi \rightarrow \pm \mathrm{i} \infty$,  making the orbital coefficients non-normalisable and the corresponding states unphysical.
As a result, the Fock operator becomes ill-defined, leading to a singularity in the energy.

\textit{The hidden connection.---}
Having revealed a deeper \alter{connectivity} between multiple HF solutions, the final key observation is that the sb-UHF solution is a ground-state wave function in the repulsive regime, but becomes an excited-state wave function for the attractive regime, as shown in Fig.~\ref{fig:energy}. 
This can be confirmed by looking at the number of nodes of the wave function.
\alter{In principle}, therefore, by slowly varying the e-e interaction strength in a similar manner to an adiabatic connection in DFT, \cite{Langreth_1975, Gunnarsson_1976} one can ``morph'' a ground-state wave function into an excited-state wave function \alter{\textit{via} a stationary path of HF solutions.}
Clearly, any path connecting the sb-UHF states of the repulsive regime to those in the attractive regime must avoid the singularity at $\lambda = 0$.
One possibility would be to follow a route ``the other way around'' the real number line, passing from $\lambda > 0 $, through $\lambda = + \infty$, and returning via $\lambda = - \infty$ at $\lambda < 0$.
Such a route would, however, involve its own obvious computational complications. 
\alter{Alternatively, one can simply} follow a complex contour around the branch cut running between the repulsive and attractive Coulson--Fischer points, as shown by the solid blue curves in Fig.~\ref{fig:riemann}. 
In such a way, one can ensure a smooth transition of the wave function coefficients from the repulsive to the attractive states \alter{whilst maintaining stationarity with respect to the parameterised Hamiltonian}.
Somehow, because one cannot order complex energies, ground and excited states are able to mix away from the real axis.

The complex adiabatic path followed to obtain the physical transition $s^2 \rightarrow p_z^2$ (at $\lambda = 1$) is shown in blue in Fig.~\ref{fig:adia}.
Starting on the RHF ground-state wave function at $\lambda = 1$, one increases $\lambda$ in order to reach the repulsive Coulson--Fischer point at $\lambda = 3/2$, where one can transfer directly to the sb-UHF state.
From this point, one follows the complex contour represented in Fig.~\ref{fig:riemann} in order to avoid the singularity at $\lambda = 0$ and the branch cut running along the real axis (see also right panel of Fig.~\ref{fig:adia}).
In doing so, one ends up on the excited sb-UHF state.
By increasing $\lambda$ again, one reaches the attractive Coulson--Fischer point at $\lambda = -75/62$, where one can transfer directly from the sb-UHF state to the $p_z^2$ RHF state. 
From here, adiabatically following the $p_z^2$ RHF state up to $\lambda = 1$ completes the complex adiabatic connection path.
\alter{Notably, in contrast to the usual density-fixed adiabatic path in DFT, we allow the HF density to relax at each $\lambda$ in a similar manner to Ref.~\onlinecite{Seidl_2018}.
The present methodology is implemented in a modified version of Q-Chem. \cite{QCHEM4}
}

\textit{Concluding remarks.---}
\alter{The use of non-Hermitian Hamiltonians as a tool for understanding stationary states in electronic structure methods is in its infancy, and many exciting properties remain to be found and understood.
Here we have presented a first study of non-Hermitian quantum mechanics for the exploration of multiple solutions at the HF level.
Albeit simple, the present model system perfectly illustrates the deeper topology of the multiple electronic states revealed using a complex-scaled e-e interaction.
Indeed, we have found identical connections in various other systems, including the Hubbard model and simple diatomics such as \ce{H2}.
In this more complex landscape, solutions are connected as part of a continuous structure of stationary states, and Coulson--Fischer points show a close resemblance to non-Hermitian exceptional-point degeneracies.
Through the introduction of non-Hermiticity, we are provided with a more general arena in which the complex and diverse characteristics of multiple solutions can be explored and understood.}

\alter{
The practical implications of non-Hermitian analytic continuations remain very much unexplored.
In the current work, we have used the construction of a complex adiabatic connection between ground and excited HF solutions as a simple first application for the determination of excited states.
Indeed, the natural stationary paths identified between multiple solutions may have wider relevance across quantum chemistry, for example in the development of novel DFT functionals, or for understanding the evolution of stationary states between different levels of theory.
In the future, we plan to extend our non-Hermitian approach to correlated methods, in particular the CC family of methods that are widely regarded as the gold standard of quantum chemistry.
Similar to HF, the non-linearity of CC methods yields a large manifold of solutions (including complex ones) as described in Ref.~\onlinecite{Kowalski_1998}.
An analytic continuation of the CC amplitudes --- instead of the orbital coefficients --- may therefore reveal a similar fundamental topology of multiple solutions.
}

\textit{Acknowledgements.---}
H.G.A.B.~thanks the Cambridge and Commonwealth Trust for a studentship and A.J.W.T.~thanks the Royal Society for a University Research Fellowship (UF110161). 
P.F.L~thanks Julien Toulouse for enlightening discussions at the early stage of this work.

%

\end{document}